\documentclass[pdflatex,sn-mathphys-num]{sn-jnl}

\usepackage{graphicx}%
\usepackage{multirow}%
\usepackage{amsmath,amssymb,amsfonts}%
\usepackage{amsthm}%
\usepackage{bbm}
\usepackage{mathrsfs}%
\usepackage[title]{appendix}%
\usepackage{xcolor}%
\usepackage{textcomp}%
\usepackage{manyfoot}%
\usepackage{booktabs}%
\usepackage{algorithm}%
\usepackage{algorithmicx}%
\usepackage{algpseudocode}%
\usepackage{listings}%
\usepackage[usestackEOL]{stackengine}
\usepackage{longtable}
\usepackage{float} 

\theoremstyle{thmstyleone}%
\theoremstyle{thmstyletwo}%

\theoremstyle{thmstylethree}%

\begin{document}

\title[Article Title]{Journal Publications in Medicine: Ranking vs. Interdisciplinarity}


\author*[1]{\fnm{Anbang} \sur{Du}}\email{ad1u21@soton.ac.uk}

\author[2]{\fnm{Michael} \sur{Head}}\email{m.head@soton.ac.uk}

\author[1]{\fnm{Markus} \sur{Brede}}\email{markus.brede@soton.ac.uk}

\affil*[1]{\orgdiv{School of Electronic and Computer Science}, \orgname{University of Southampton}, \orgaddress{\city{Southampton}, \postcode{SO17 1BJ}, \country{UK}}}

\affil[2]{\orgdiv{Faculty of Medicine}, \orgname{University of Southampton}, \orgaddress{\city{Southampton}, \postcode{SO17 1BJ}, \country{UK}}}


\abstract{Interdisciplinary research is critical for innovation and addressing complex societal issues. We characterise the interdisciplinary knowledge structure of PubMed research articles in medicine as correlation networks of medical concepts and compare the interdisciplinarity of articles between high-ranking (impactful) and less high-ranking (less impactful) medical journals. We found that impactful medical journals tend to publish research that are less interdisciplinary than less impactful journals. Observing that they bridge distant knowledge clusters in the networks, we find that cancer-related research can be seen as one of the main drivers of interdisciplinarity in medical science. Using signed difference networks, we also investigate the clustering of deviations between high and low impact journal correlation networks. We generally find a mild tendency for strong link differences to be adjacent. Furthermore, we find topic clusters of deviations that shift over time. In contrast, topic clusters in the original networks are static over time and can be seen as the core knowledge structure in medicine. Overall, journals and policymakers should encourage initiatives to accommodate interdisciplinarity within the existing infrastructures to maximise the potential patient benefits from IDR.}




\maketitle

\section{Introduction}\label{sec1}
Interdisciplinary research (IDR) is generally believed to be an important source of creativity and innovativeness \cite{NAP11153,rafolsDiversityNetworkCoherence2010,Rousseau_knowledge_2019}. According to the National Academies of Sciences of the USA \cite{NAP11153}, IDR is "a mode of research by teams or individuals that integrates information, data, techniques, tools, perspectives, concepts and/or theories from two or more disciplines or bodies of specialized knowledge to advance fundamental understanding or to solve problems whose solutions are beyond the scope of a single discipline or area of research practice." 

The relationship between interdisciplinarity and impact has been a longstanding topic of the science of science and science policy communities, with core research questions evolving around how IDR impacts science and thus many of the complex research problems human society is currently facing \cite{chenAreTopcitedPapers2015,okamuraInterdisciplinarityRevisitedEvidence2019a,molas-gallartRelationshipInterdisciplinarityImpact2014,lariviereRelationshipInterdisciplinarityScientific2010,yegros-yegrosDoesInterdisciplinaryResearch2015a,caiRelationshipInterdisciplinarityCitation2023,huInterdisciplinaryResearchAttracts2024b}. 

Some work has suggested that IDR tends to more impactful in terms of citations \cite{chenAreTopcitedPapers2015,okamuraInterdisciplinarityRevisitedEvidence2019a} while others argue against this because the way how interdisciplinarity and citations are measured could influence the results \cite{molas-gallartRelationshipInterdisciplinarityImpact2014,lariviereRelationshipInterdisciplinarityScientific2010,yegros-yegrosDoesInterdisciplinaryResearch2015a}. The authors of \cite{caiRelationshipInterdisciplinarityCitation2023} have revealed a more nuanced relationship between scholarly impact and interdisciplinarity: the more interdisciplinary papers tend to have delayed recognition and greater long-term citation sustainability compared with more disciplinary work. Beyond scholarly impact, interdisciplinary research has also been found to attract more policy attention \cite{huInterdisciplinaryResearchAttracts2024b}.


One noted issue regarding IDR is the conflict between strong advocacy at policy level \cite{woelertParadoxInterdisciplinarityAustralian2013, wagnerApproachesUnderstandingMeasuring2011a} and poor reward at research evaluation due to rigid disciplinary-based standards \cite{bromhamInterdisciplinaryResearchHas2016a,rafolsHowJournalRankings2012}, i.e., the “paradox of interdisciplinarity”  \cite{woelertParadoxInterdisciplinarityAustralian2013}. 

Scientific journals are an important medium for scholarly communications \cite{milojevicNatureSciencePNAS2020,andresMedicalJournalsRole2024}. High-ranking journals, measured by citation impact and prestige indicators, are at the forefront of leading and shaping domain knowledge, trends, innovations, and practices \cite{milojevicNatureSciencePNAS2020,andresMedicalJournalsRole2024}. There has been limited attention in understanding the role of high-ranking journals in disseminating interdisciplinary knowledge and practices. In this paper, along with the previous work \cite{du_prestigious_2025_frccs}, we ask if high-ranking (highly impactful hereafter) journals lead in disseminating IDR, or tend to be more disciplinary-based. 

As relevant work in the past has provided limited evidence in medicine, we will explore this question specifically for medical research. This is because pressing health issues make medical research highly consequential and the direct population benefits that medical research generates mean that understanding how journals shape knowledge flows has practical implications for patient outcomes and public health \cite{Du2025lancet}. Furthermore, knowledge integration across medical areas has demonstrated great benefits to patients, medical science and society \cite{schwetzExtendedImpactHuman2019a}, yet the role of journal prestige in facilitating or discouraging such integration remains unclear. Comparing high and low  impact  journals, this study provides specific insights into how medical journals can better support interdisciplinary work.

Characterising the interdisciplinary knowledge structure of PubMed research articles in medicine as correlation networks of medical concepts and comparing the interdisciplinarity of articles from highly impactful journals (I) and less/not impactful journals (NI), we aim to answer the following research questions:
\begin{enumerate}
    \item Do I journals produce more interdisciplinary work than NI journals in medicine? 
    \item Which topic areas drive the development of IDR in I and NI journals?  
    \item Do large differences in correlation networks for I and NI occur in places of otherwise strong/weak correlations? 
    \item Do differences between correlation networks in I and NI cluster around certain topic areas? If so, which topic areas are they?  
\end{enumerate}

The novelty of this work, compared with the previous conference paper \cite{du_prestigious_2025_frccs}, is threefold. Firstly, while the analysis of differences between I and NI networks in \cite{du_prestigious_2025_frccs} was done through the lens of absolute difference networks, by constructing and analysing positive and negative difference networks in Section \ref{sec: Differences between I and NI-June Networks} this work takes the signed (positive and negative) differences into consideration. This includes the analysis of signed link co-location in Section \ref{sec: Do strong differences co-locate?} and clustering in Section \ref{sec: Do differences cluster?}. Secondly, to better understand network connectivity and topic area prominence, we introduce a new analysis of inner cores through network decomposition via link thresholding, done for both I and NI networks in Section \ref{sec: Core Analysis} and signed difference networks in Section \ref{sec: Do differences cluster?}. Thirdly, in Section \ref{sec: Do differences cluster?} we also analyse the relationship between the link strength of the signed difference networks and the original networks.




\section{Method}
\label{Method}
In this paper, we are interested in comparing IDR in impactful and less impactful medical journals. We thus define a journal to be impactful in a given year if it was in the top 10\% of the SCImago Journal Rank (SJR)\footnote{\url{https://www.elsevier.com/en-gb/products/scopus/metrics}} in medicine in that year, where the SJR is a widely used metric that weighs the value of a citation based on the subject field, quality and reputation of the source.\footnote{Traditional journal ranking like the Journal Citation Report (JCR) from the Web of Science reports field-specific quantile ranking, where the top 25\% (quantile 1) is generally deemed as high-impact journals in the field. We impose a more strict 10\% cutoff as our definition of highly impactful journal here.}

\begin{figure}[htp]
    \centering
    \includegraphics[width=0.5\linewidth]{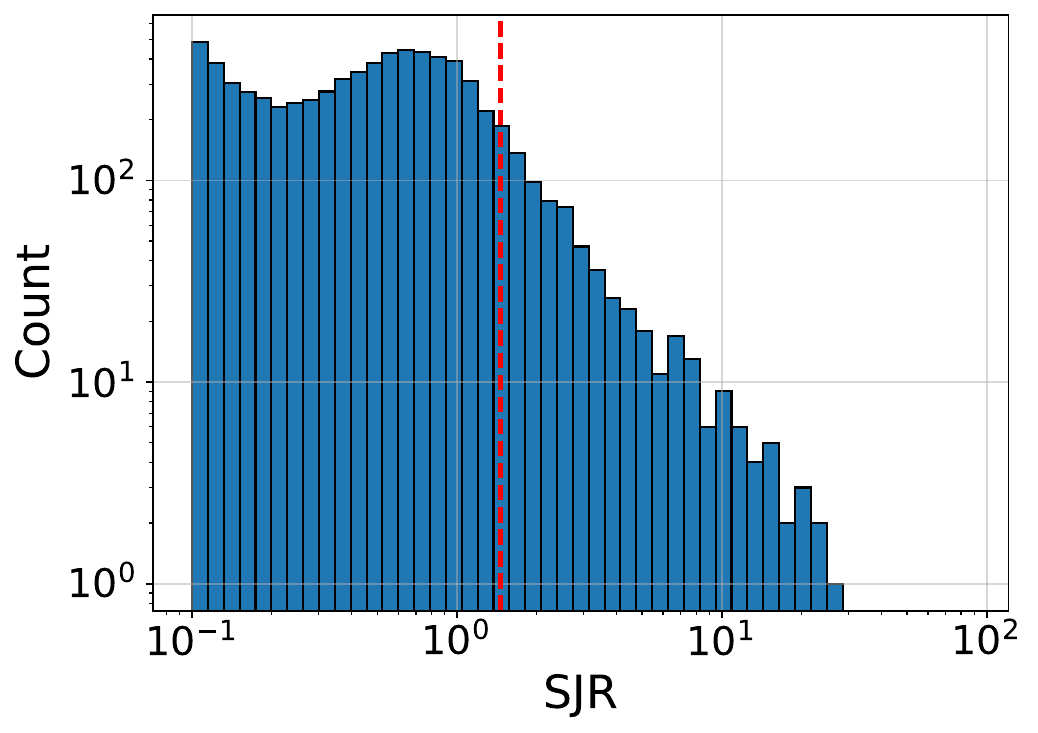}
    \caption{Distribution of impacts (SJR) of the Scimago journal ranking for the year 2022 in medicine. Red vertical line (SJR=1.46) represents the 10\% cutoff that captures 42\% of total impact of 7187 journals.}
    \label{fig:sjr_distribution}
\end{figure}

The reason for this 10\% cutoff is twofold. First, journal rankings outside the top tier journals are often instable \cite{pajic2015stability,polonsky2005we}, boundaries between traditional impact quartiles often lack robustness, i.e., their rankings tend to fluctuate over time. Second, as shown in Fig.\ref{fig:sjr_distribution}, the distribution of impacts (SJR) of the Scimago journal ranking (2022) has a long-tailed distribution, meaning that only few journals are exceptionally impactful while a vast majority of journals has impact that is confined to a narrow range of small values. The 10\% cutoff coincides with the start of the tail. Through picking the 10\% cutoff (red vertical line, SJR=1.46, 42\% of total impact captured), we partition the journal space into two groups of clearly distinct journal impact.

We characterise the interdisciplinary knowledge structure by correlation networks of medical concepts, i.e., Medical Subject Heading (MeSH)\footnote{\url{https://www.nlm.nih.gov/mesh/meshhome.html}}, indexed in research articles in PubMed. MeSH, as a hierarchical collection of medical concepts curated and maintained by the National Library of Medicine of the US, and is one of the most comprehensive taxonomies of medical research \cite{leblancAddedValueMedical2024}. Each PubMed article has been indexed with a number of MeSH terms representing its thematic focus. In this paper, to maintain balance between granularity and computational feasibility of data gathering, we focused on the second-level MeSH terms under the biggest branch ``Disease" (C-branch). This branch contains communicable and non-communicable diseases (for example, HIV, tuberculosis, stroke and diabetes) which are persistently major global health issues \cite{mcintoshGlobalFundingCancer2023,headAllocationUS105Billion2020,Du2025lancet}. Knowledge integration across diseases, and therefore interdisciplinarity,  has demonstrated great benefits to patients, medical science, and society \cite{duIntegrationVsSegregation2025a,schwetzExtendedImpactHuman2019a}.

Following the approach in earlier related  paper on the topic \cite{duIntegrationVsSegregation2025a,du_prestigious_2025_frccs}, we construct co-occurrence networks of C-branch second-level MeSH terms for research articles published in impactful (I) and less-impactful (NI) journals respectively in 1999, 2010, and 2022. Note, that we only selected three time points to study temporal change. This choice was motivated by pragmatic considerations in data collections and rate limits of the PubMed API. Specifically, we started data collection for 1999, as this is the first year in which SJR medical rankings were made available\footnote{\url{https://www.scimagojr.com/journalrank.php}}. 2010 and 2022 were selected as a midpoint and endpoint that are both separated by a decade, allowing to capture the time-dependence of the networks over longer time scales.

The co-occurrence count $c_{ij}$ between MeSH term $i$ and $j$ records the number of research articles that have indexed both terms. Further following \cite{duIntegrationVsSegregation2025a,du_prestigious_2025_frccs,eckHowNormalizeCooccurrence2009b}, the co-occurrences were then normalised through the cosine similarity to reflect the extent of knowledge integration between concepts $i$ and $j$, i.e., $w_{ij}=c_{ij}/\sqrt{c_{ii}c_{jj}}$, where $c_{ii}$ and $c_{jj}$ measure the number of articles indexed with term $i$ and $j$ respectively. A high value of $w_{ij}$ thus implies a strong correlation between the respective pair of MeSH terms, i.e., represents a pair of well-integrated concepts. 

Note that in this paper, we adopt a network model of interdisciplinarity that is operationalised at a higher level of aggregation (each impact tier is treated as a collection of articles) than at the individual article level. The resulting model is computationally easier to construct as one does not have to download and process millions of relevant publications. Alternatively, one could have used measures of interdisciplinarity at individual article level, e.g. diversity \cite{rafolsDiversityNetworkCoherence2010}. Future work might aim to gain further insights using such measures.

Data collection and construction of the correlation network resulted in $296$ nodes (second-level MeSH terms) for all three years. Preliminary analysis showed that the resulting networks were not always connected. As we are interested in analysis of knowledge integration over the entire complex system of medical research, we excluded MeSH terms that correspond to isolated nodes over all points in time. Following this procedure, we obtain a core network composed of $201$ nodes for all three years, which we use as the basis of this study below.

As preliminary analysis also indicated the possibility of sample size effects due to the much larger number of papers used to construct the NI network, we also constructed a version of the NI network based on a smaller number of papers. This was done by restricting paper collection to only one month in the year when collecting NI data. After initial comparisons, which revealed no particular biases depending on the choice of month, without losing generality, we picked the month of June, i.e., the midpoint of a year, and labelled the corresponding data NI-June. 

To study the dependency of component decomposition with respect to the link weight threshold in Fig.\ref{LCC}(A) and Fig.\ref{core pos neg}(A), we generate 100 reference networks for each empirical network through random link rewiring across all pairs of nodes. For this purpose, randomly selected pairs of nodes were chosen and their respective links swapped. The procedure was iterated 100 times. In the figures, blue lines indicate the decomposition of the empirical networks and red lines the mean of the $100$ randomised networks with red ribbons indicating 95\% confidence intervals. 

To explore thematic differences between IDR in impactful journals and less impactful journals, we also subtracted link weights of NI-June networks from I networks, i.e., $w_{ij}^I-w_{ij}^{NI-june}$, and constructed the positive and negative difference network separately for comparison. Note that in Fig. \ref{fig: diff vs I}, the y axis represents the percentage deviation from the mean link strength of the I network of that year. We introduce this standardisation to adjust for potential bias across time.

In this study, we adopt several network measures to quantify different aspects of interdisciplinarity. Following \cite{rafolsDiversityNetworkCoherence2010}, we define the mean node strength to represent on average how concepts integrate with the neighbouring concepts, i.e.,
\begin{equation}
    \overline{s} = \frac{1}{N}\sum_is_i = \frac{1}{N}\sum_i\sum_{j\neq i}w_{ij}
\end{equation}
where $N$ represents the number of nodes in the network. $\overline{s}$ is often called network coherence in relevant literature \cite{rafolsDiversityNetworkCoherence2010}. Note that the node strength $s_i$ \cite{barratArchitectureComplexWeighted2004} is a measure of how well a concept integrates knowledge locally.

We propose the average shortest path length (ASPL) \cite{jahanshadGeneticsPathLengths2012} to capture the average shortest distance between the concepts. A low ASPL indicates that the disease network is compact. The shortest path length \cite{brandes_faster_2001} between node $i$ and $j$ is 
\begin{equation}
    d(i,j) = \min\left(\frac{1}{w_{ih}}+\cdots+\frac{1}{w_{hj}}\right)
\end{equation}
with $h$ being intermediary nodes between node $i$ and $j$. ASPL is the average of $d(i,j)$ over all possible pairs of concepts $(i,j)$.

Following \cite{leydesdorff_betweenness_2007}, betweenness centrality (BC) \citep{brandes_faster_2001} is used to track node-level interdisciplinarity from the intermediation perspective, i.e., one concept's ability to bridge otherwise disjoint disease knowledge. BC of a node $i$ is computed as 
\begin{equation}
    b_i = \sum_{j\neq k}\frac{g_{jk}(i)}{g_{jk}}
\end{equation}
where $g_{jk}$ is the number of shortest path between any two nodes and $g_{jk}(i)$ is the number of shortest path between two nodes going through node $i$. 

Exploring clustering patterns of the medical concepts, we evaluated the networks modularity and used the Louvain method to determine modules of interconnected concepts \cite{blondelFastUnfoldingCommunities2008}. Modularity is computed as 
\begin{equation}
    Q = \frac{1}{2m}\sum_{i,j}\left(w_{ij}-\frac{s_is_j}{2m}\right)\delta(c_i,c_j)
\end{equation}
where $m$ is the sum of link weights, $s_i$ refers to the strength of node $i$, and $\frac{s_is_j}{2m}$ is the expected link strength between $i$ and $j$ assuming a random distribution of connections which preserves the strength distribution across nodes. A high modularity $Q$ represents a well-defined community structure with many intra-community links and few links connecting separate communities, while a low $Q$ indicates a weak community structure.

We also introduce global clustering coefficient ($GCC$) 
\cite{opsahlClusteringWeightedNetworks2009}, which is computed as:
\begin{equation}
GCC =\frac{\sum_j\sum_{\{i,k\}\subset N(j)}w_{ijk}A_{ik}}{\sum_j\sum_{\{i,k\}\subset N(j)}w_{ijk}}
\end{equation}
where $N(j)$ is the set of neighbouring nodes of $j$ and $w_{ijk}=(w_{ij}+w_{jk})/2$. $A_{ik}=1$ if $i$ is connected with $k$, otherwise $A_{ik}=0$. $GCC$ measures the proportion of the weights of closed triads out of open triads. A high $GCC$ indicates a strong triadic closure within the interdisciplinary knowledge network, i.e., two concepts sharing a common neighbouring concepts are more likely to be co-studied.

Following \cite{newman2002,limDiscordantAttributesStructural2019}, we adopted the measure of node strength assortativity $r$ to exploring the mixing pattern of the interdisciplinary knowledge network. Let $G=(V,E)$ be an undirected graph with set of vertices $V$ and undirected edges $E$, we define reciprocal edge set $\mathcal{E}:=\{(i,j):\{i,j\}\in E\}\cup\{(j,i):\{i,j\}\in E\}$ and $\mu:=\frac{1}{|\mathcal{E}|}\sum_{(i,j)\in\mathcal{E}} s_i$. The node strength assortativity $r$ is the Pearson's correlation across all edges in the reciprocal edge set $\mathcal{E}$:
\begin{equation}
r_s \;=\;
\frac{\sum_{(i,j)\in\mathcal{E}} (s_i-\mu)(s_j-\mu)}{\sum_{(i,j)\in\mathcal{E}} (s_i-\mu)^2 }=\frac{cov(s_i,s_j)}{var(s_i)}
\end{equation}
Ranging between $[-1,1]$, a high positive $r$ indicates a widely-connected concept tends to be co-studied with other widely-connected concept, i.e., assortative mixing. A large negative $r$ indicates a widely-connected concept tends to be co-studied with other less-connected concept, i.e., disassortative mixing. $r\approx0$ indicates a rather random mixing pattern.

To study RQ1, we adopt mean node strength to measure and compare the interdisciplinarity between impactful and non-impactful networks. As regards RQ2, we study the rankings of node strength and betweenness of different topic areas. To address RQ3, we first construct the difference network (described above) and conduct regression analysis of link weight on adjacent link weights and use this to evaluate the tendency of co-location of strong/weak differences through measuring the regression slope. Finally, to study RQ4, we perform community decomposition through the Louvain algorithm and analyse community membership.

\section{Results}\label{sec2}
In this section, to answer our research questions, we conduct experiments and analyse the results. To gain intuition, in Section \ref{sec: Knowledge Network and Clusters}, we visualise the interdisciplinary knowledge network and the clusters. Then, to allow proper comparison between impactful and less impactful IDR, we analyse the networks' link and node strength distributions and discuss a sample size effect in network construction and how to correct for it in Section \ref{sec: Sample Size Effect}. 

To gain insights into different aspects of interdisciplinarity related to impactful and less impactful IDR, we compute and compare global network properties of the corresponding networks in Section \ref{sec: Global Network Properties}. As we are also interested in link placement, we explore the relative arrangement of strong and weak links in the networks in Section \ref{sec: Core Analysis}. Furthermore, to better understand interdisciplinarity at a more granular level, in Section \ref{sec: Node and Community Importance}, we study node-level and community-level importance. Finally, in Section \ref{sec: Differences between I and NI-June Networks}, we study if the differences between I and NI-June networks cluster around particular topic areas. 

\subsection{Knowledge Network and Clusters}
\label{sec: Knowledge Network and Clusters}
\begin{figure}
    \centering
    \includegraphics[width=\linewidth]{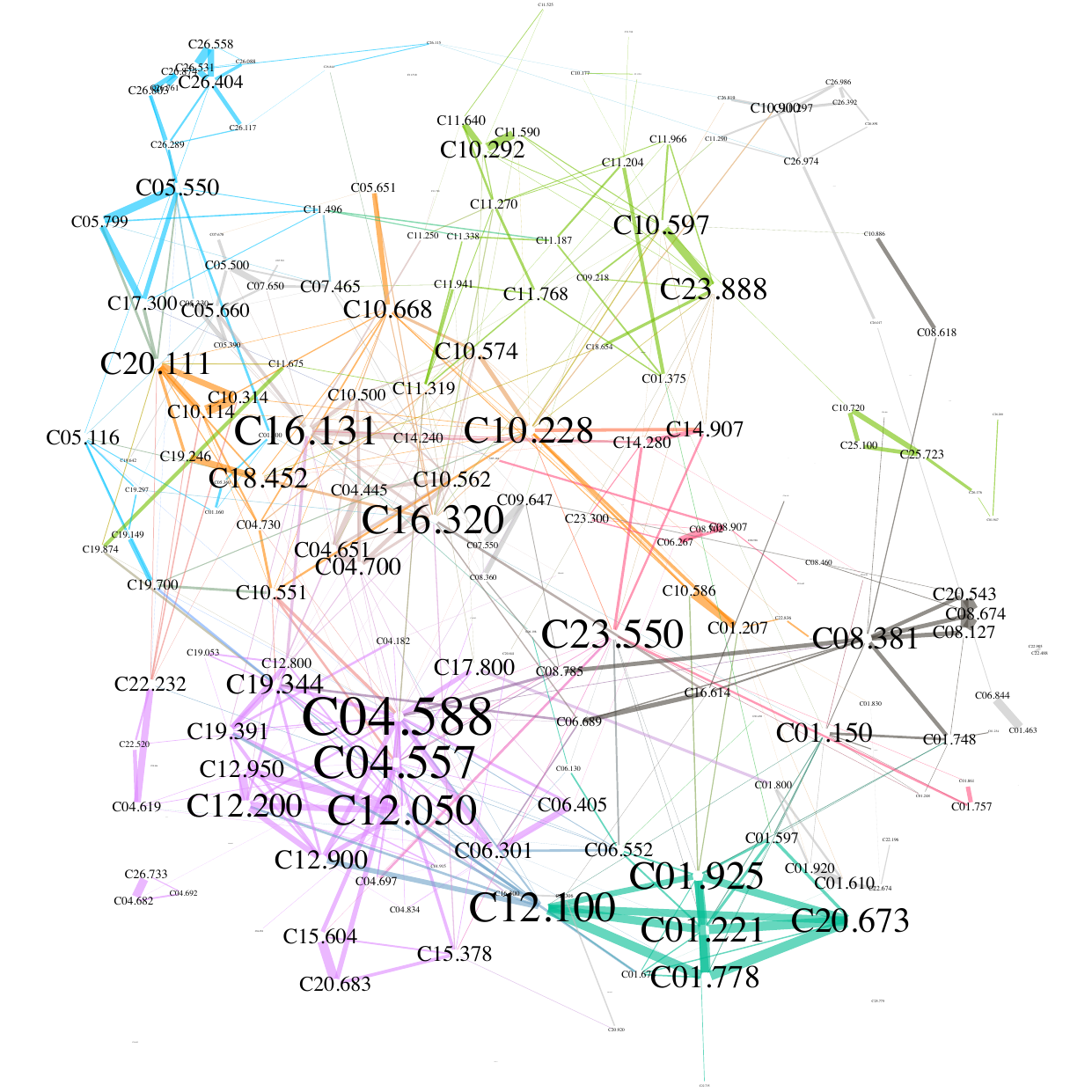}
    \caption{Illustration of the Impactful network in 1999. Links with strength less than $0.080$ were filtered out for better visualisation. Communities were detected based on the Louvain algorithm \cite{blondelFastUnfoldingCommunities2008} and are labelled by MeSH codes. Label size represents the size of node strength. We find a very similar community breakdown for the non-impactful network in 1999 (not shown).}
    \label{fig:1999I}
\end{figure}

To gain intuition, we start by visualising the impactful network in 1999, see Fig. \ref{fig:1999I}, where nodes are coloured by topic clusters following modularity decomposition using the Louvain method \cite{blondelFastUnfoldingCommunities2008}. As one might expect, we observe significant modularity ($Q=0.47$) and a breakdown into several major clusters that align with medical topics, including a cancer-related cluster (C04 pink), an infectious diseases related cluster (C01 cold green), a nervous system diseases related cluster (C10 warm green), a respiratory diseases related cluster (C08 dark grey), and a musculoskeletal diseases related cluster (C05 blue). C16 (Congenital, Hereditary and Neonatal Diseases and Abnormalities) and C23 (Pathological Conditions, Signs and Symptoms) appear to be intermediaries that connect different clusters. 

\subsection{Sample Size Effect}
\label{sec: Sample Size Effect}
\begin{figure}
    \centering
    \includegraphics[width=\linewidth]{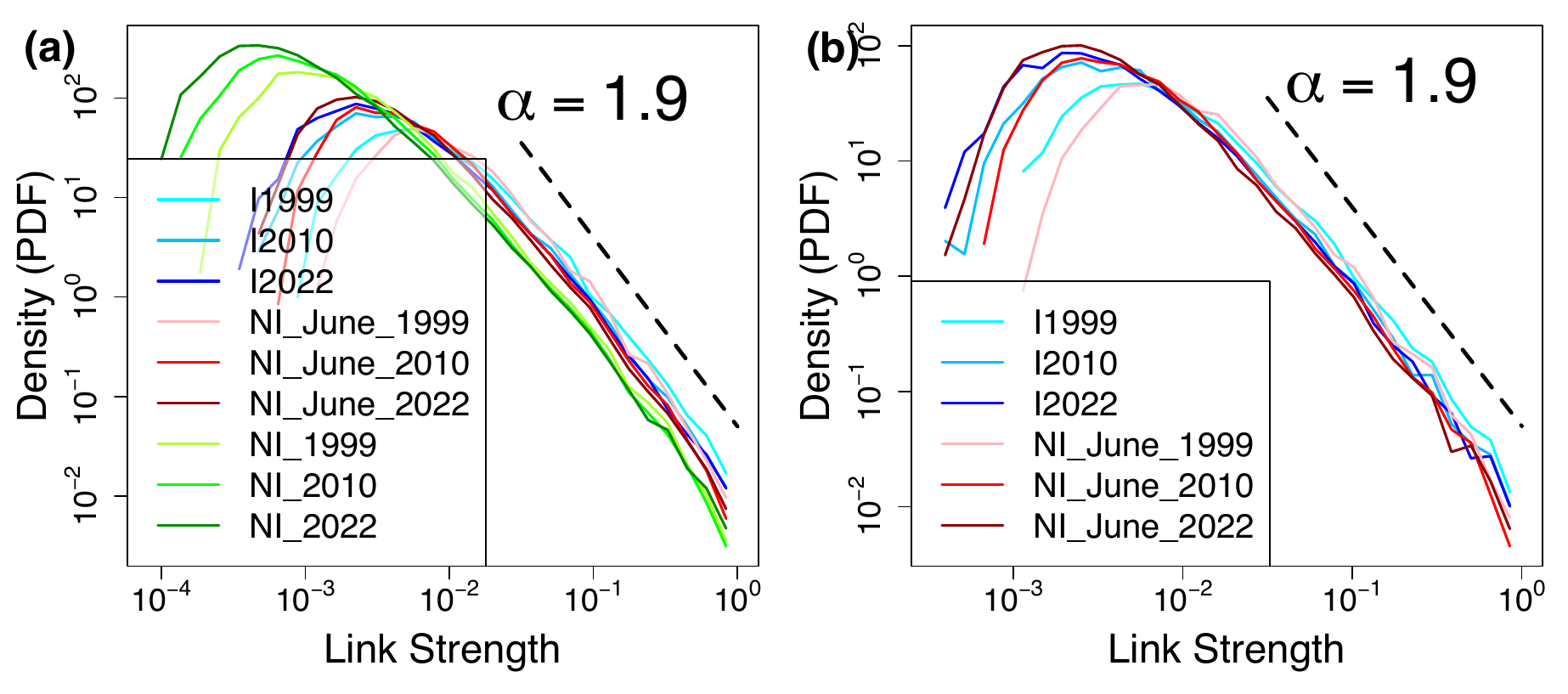}
    \caption{Link Strength Distribution for (a) the I, NI, and NI-June networks, and (b) the I and NI-June networks. Logarithmic binning was applied on the x-axis with $30$ bins. For comparison, a power law distribution with exponent $\alpha=1.9$ and minimum value of link strength $x_{min}=10^{-1.5}$ is plotted as the dotted line against the tails of the distributions.}
    \label{fig:link_strength}
\end{figure}

Plotting the link strength distribution for I and NI networks for 1999, 2010 and 2022 in Fig. \ref{fig:link_strength}), we first note that the distributions show approximately power-law distributions over several orders of magnitude in the x and y dimensions with a power law exponent close to $2$. Next, we observe that the NI networks tend to have a higher frequency of weaker links and less frequently stronger links than I (Fig. \ref{fig:link_strength}.a). We hypothesize that this may be due to a sample size effect, as the network constructed for NI is based on a much larger set of papers than the I network Table \ref{tab:paper number}. Adjusting for sample size, we note that the dominance of weak links disappears when comparing approximately equal samples (Fig. \ref{fig:link_strength}.b, Table \ref{tab:paper number}), i.e. the data for I and the NI sample collected for the month of June, which supports our hypothesis.


We further note that for all samples, I, NI, and NI-June, the number of papers published has increased strongly over time, leading to sample size differences between the different snapshots in time, see Table \ref{tab:paper number}. This makes it difficult to compare different snapshots in time and we leave the temporal analysis for future work and focus on the comparison between I and NI-June networks in this paper.

\begin{table}
    \centering
    \begin{tabular}{cccc}
    \hline
    Number of Papers  & 1999 & 2010 & 2022 \\
    \hline
    I & 13411 & 36507 & 56421\\
    \hline
    NI June & 16051 & 39971 & 68345 \\    
    \hline
    NI & 219827 & 404115 & 663182 \\   
    \hline
    \end{tabular}
    \caption{Total number of journal papers used to construct the networks. }
    \label{tab:paper number}
\end{table}

\begin{figure}
    \centering
    \includegraphics[width=0.5\linewidth]{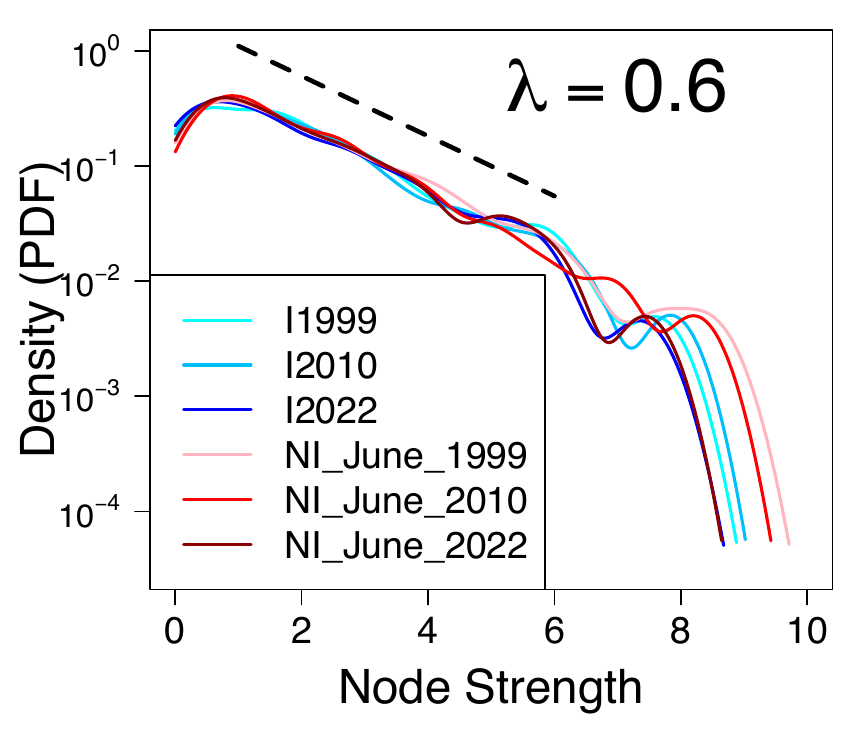}
    \caption{Node strength distribution of the I and NI networks. An exponential distribution with decay rate $\lambda=0.6$ with node strength $s\in[1,6]$ is plotted as the dotted line against the distributions.}
    \label{fig:degree}
\end{figure}

Next, we also plotted the node strength distribution for the I and NI-June networks. We note that the distributions shows approximately an exponential distribution for node strengths in the range between $1$ and $6$, see Fig. \ref{fig:degree}.

\subsection{Global Network Properties}
\label{sec: Global Network Properties}
As noted in section \ref{Method}, we compute several global network statistics in Table \ref{tab: Global Net Measures} to reveal different aspects of interdisciplinarity in medical science. First, we explore the coherence of the networks \cite{rafolsDiversityNetworkCoherence2010} which can be quantified through how well an average node connects with its neighbouring nodes, i.e. the average node strength $\overline{s}$. We observe that with $\overline{s}_{NI,June}=1.85, 1.81$ and $1.73$ (for the three time periods)  the NI-June network has a higher average node strength than the I network with $\overline{s}_I=1.75, 1.71$ and $1.64$. This indicates that non-impactful research is more interdisciplinary than impactful research. The result is also supported by the finding that the NI-June networks have consistently higher network density than the I networks, i.e., there are consistently more and stronger connections between medical concepts in non-impactful work than in the more impactful work. From the computed network density, we also note that the networks are not sparse.

Another important aspect of interdisciplinarity is the compactness of the knowledge networks. In bibliometrics analyses of co-occurence networks this is often measured by the average shortest path length $ASPL$ \cite{rafolsDiversityNetworkCoherence2010,duIntegrationVsSegregation2025a,jahanshadGeneticsPathLengths2012}. In Table \ref{tab: Global Net Measures} we also observe that $ASPL$ of the NI-June networks is consistently slightly higher than for the I networks, i.e., we find a slightly higher level of compactness of NI-June networks. We conclude that knowledge flows between medical concepts are somewhat more supported in NI journals. 

Exploring global network clustering patterns, we compute the modularity $Q$ \cite{blondelFastUnfoldingCommunities2008} to quantify the strength of clusters and the global clustering coefficient $GCC$ \cite{opsahlClusteringWeightedNetworks2009} to capture the tendency for triadic closure. From Table \ref{tab: Global Net Measures}, we observe that the I network exhibits higher network modularity $Q$ than the NI-June network, despite that this effect becomes less pronounced in later years. This suggests that impactful interdisciplinary research tends to be more compartmentalised, i.e., focuses on smaller coherent knowledge clusters. In terms of global clustering coefficient $GCC$, we find no significant differences between the I and the NI-June network. Taking the observations of $Q$ and $GCC$ together, because modularity values differ substantially while clustering coefficients are similar, we note that for the I networks the modular organisation is at a larger scale than the tendency of triadic closure would suggest.

Exploring mixing patterns of bodies of knowledge, we also compute the node strength assortativity $r$ \cite{newman2002,limDiscordantAttributesStructural2019}. From Table \ref{tab: Global Net Measures}, we observe that the node strength assortativity $r$ for all networks tend to be moderately negative with no significant differences in magnitude between the I and the NI-June network. This indicates that widely-connected medical concepts tend to be preferentially co-studied with other less-connected concept. Our observation of negative assortativity $r$ is also consistent with other studies of MeSH networks \cite{kastrinLargeScaleStructureNetwork2014} and semantic networks \cite{budelTopologicalPropertiesOrganizing2023} that are both hierarchical concept networks.

\begin{table}
    \centering
    \begin{tabular}{cccc}
    \hline
     Global Network Measure  & 1999 & 2010 & 2022 \\
    \hline
     $\overline{s}_I$   & 1.75 & 1.71  & 1.64\\
     $\overline{s}_{NI,June}$ & 1.85 & 1.81 & 1.73\\
     
    \hline
    Edge Density $I$  & 0.18 & 0.24 & 0.24 \\
    Edge Density $NI$ & 0.22 & 0.29 & 0.31 \\
    \hline
    $Q_I$ & 0.47 & 0.45 & 0.45 \\
    $Q_{NI,June}$ & 0.41 & 0.41 & 0.42 \\
    \hline
    $GCC_I$  & 0.59 & 0.63 & 0.67 \\
    $GCC_{NI,June}$ & 0.58 & 0.64 & 0.68 \\
    \hline
     $ASPL_I$   & 26.28 & 26.52  & 29.54\\
     $ASPL_{NI,June}$ & 24.38 & 25.26 & 27.51\\
    \hline
     $r_I$   & -0.18 & -0.19  & -0.21\\
     $r_{NI,June}$& -0.19 & -0.19  & -0.21\\
     \hline
    \end{tabular}
    \caption{Average node strength $\overline{s}$ \cite{rafolsDiversityNetworkCoherence2010}, number of links $N$, modularity $Q$ \cite{blondelFastUnfoldingCommunities2008}, global clustering coefficient $GCC$ \cite{opsahlClusteringWeightedNetworks2009}, average shortest path length $ASPL$ \cite{brandesFasterAlgorithmBetweenness2001} and node strength assortativity $r$ \cite{newman2002} for the I and NI-June networks. }
    \label{tab: Global Net Measures}
    
\end{table}

\subsection{Core Analysis}
\label{sec: Core Analysis}
\begin{figure}
    \centering
    \includegraphics[width=\linewidth]{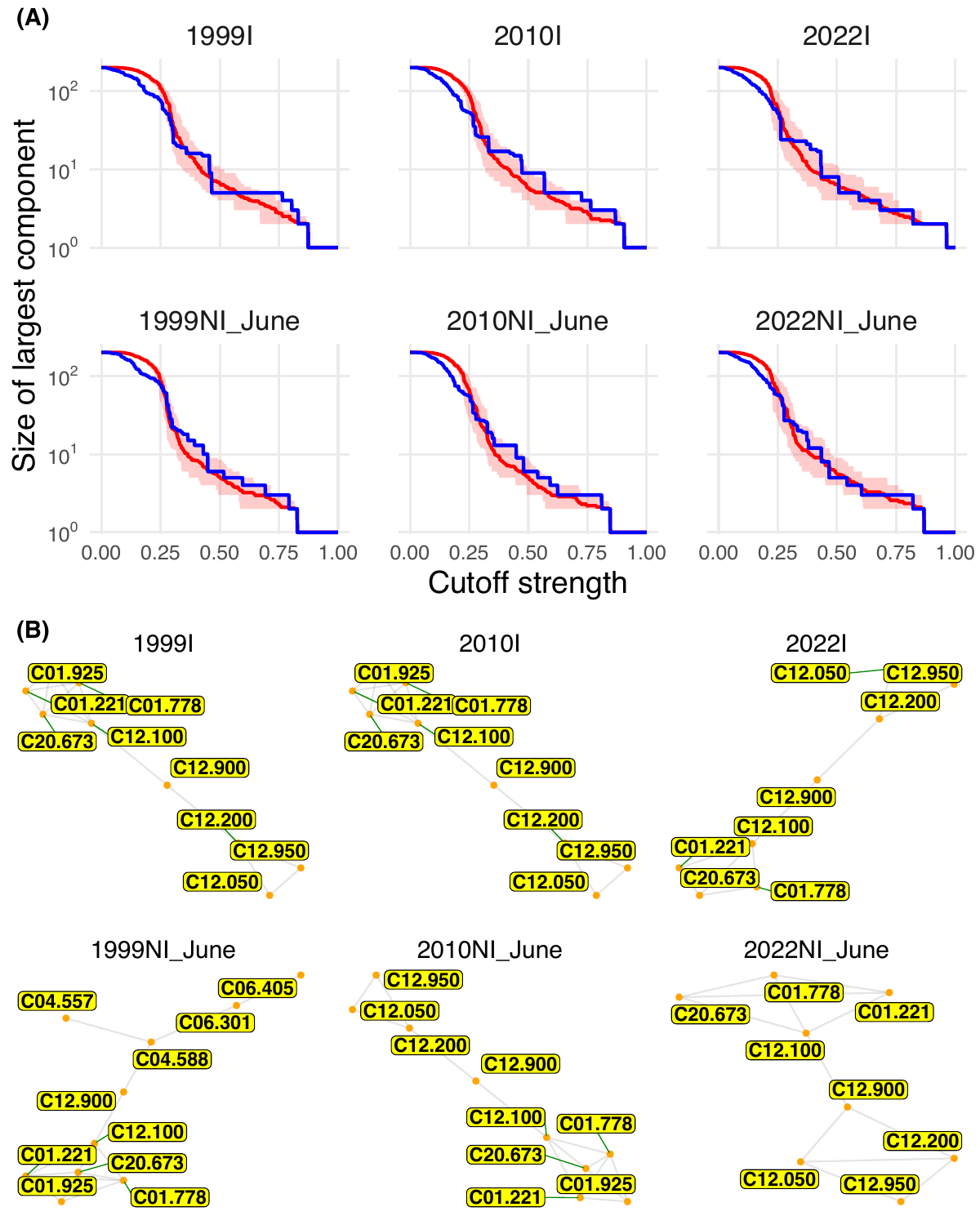}
    \caption{(A) Dependence of the size of the largest connected component (LCC) on the cutoff threshold for the I and NI June for all three points in time. The blue line represents the empirical network and the red line represents the mean of 100 randomised networks (through link shuffling) with 95\% confidence interval (red ribbon). Note that the observed cut-off of the red ribbon around $0.8$ in each network is due to the fact that the threshold reaches the largest link weight, beyond which there is no variation of the LCC size. (B) Inner core component membership (see text for a definition) for the I and NI June networks. There are cases where LCC of size 10 does not exist. In those cases LCC of size that are immediately smaller than 10 are shown, namely, size 9 for 1999I, 2010I, 2010NI-June, size 8 for 2022I, 2022NI-June, and size 10 for 1999 NI-June. }
    \label{LCC}
\end{figure}

As our correlation networks are weighted network, we are next interested in the arrangement of strong and weak links. To better understand these, we define link-strength cutoffs and analyse largest connected components of the networks composed of only links with weights that exceed the cutoff threshold. We visualise the result of the dependence of largest connected components on the cutoff threshold for the three temporal snapshots for the I and NI-June networks in Fig. \ref{LCC}(A). In the figure, blue lines indicate the decomposition of the empirical networks and red lines the mean of $100$ randomised networks (through random link rewiring mentioned in Section \ref{Method}) with red ribbons indicating 95\% confidence intervals. 

Analysing the data in Fig.\ref{LCC}(A), we observe that the LCC decomposition curve (blue) of all six networks exhibited an initially faster decay than expected at random (red) for low cut-off thresholds around 0-0.25. As this decay is gradual losing nodes one by one, this is suggestive of a core periphery structure of the empirical network such that eliminating weaker links would gradually first trim off peripheral nodes that are weakly attached to the LCC.

In Fig.\ref{LCC}(A), we find that the networks decompose from having a large LCC that comprises the entire network to smaller LCCs when link strength cutoffs are increased. We note that during this process there exists one or multiple LCCs that maintain a constant size around size $10$ or just below for a wide range of cutoffs for all the six networks, demonstrating the existence of stable inner cores  under decomposition. These inner cores of the networks are tightly connected by the strongest links and are amongst the last clusters before the network completely decays. We illustrate these inner-core memberships for each network in Fig. \ref{LCC}(B). It is notable that certain concepts occur in all of these cores, including C01.778 (Sexually Transmitted Diseases), C01.221 (Communicable Diseases), C20.673 (Immunologic Deficiency Syndromes), C12.100 (Genital Diseases), C12.900 (Urogenital Neoplasms). These concepts represent the most tightly connected topic areas and are present in the cores in all time slices as well as both in the I and NI networks.

\subsection{Node and Community Importance}
\label{sec: Node and Community Importance}
Below, we continue by analysing node importance in the networks. The corresponding statistics are reported in Table \ref{tab:local measures}, where we focus on the individual ranking of nodes in various centrality measures. As we are mostly interested in node rankings regarding overall importance and bridging effects, we focus on measuring node strength $s_i$ \cite{rafolsDiversityNetworkCoherence2010} and node betweenness $b_i$ \cite{leydesdorff_betweenness_2007,brandes_faster_2001}. 

We make the following observations. Neoplasm by site (C04.588) leads in node strength $s_i$ and node betweenness $b_i$ in both I and NI-June all the time, highlighting its crucial role in both integrating knowledge with relevant medical concepts and facilitating exchanges between distant knowledge clusters in medical research. We note that central Nervous System Diseases (C10.228) appears in the top two regarding betweenness centrality, but is not found in the top ranked nodes regarding strength. This indicates that research on CNS diseases relatively lacks of local knowledge integration with other medical concepts but serves a strong bridge connecting rather distant bodies of knowledge. 

To gain further insights into the patterns of knowledge integration, we aggregated link weights of each network at a higher level of aggregation in the MeSH hierarchy. For practical purposes and a high-level comparison, we chose the first level of category C in the MeSH hierarchy, as this only leaves us with $22$ concepts. Each first-level term could be seen as a (pre-imposed) hierarchical community consisting of a number of second-level terms. We partitioned the resulting total community-wide connection strength $s_c$ into intra-community connection strength $s_c^{intra}$ and inter-community connection strength $s_c^{inter}$. Results for all three measures are reported in Table \ref{tab:local measures}. Observing the results, it becomes clear that the ranking of $s_c$ is very stable over time, i.e., Infections (C01), Nervous System Diseases (C10) and Neoplasms (C04) consistently rank in the top three. 

High $s_c$ of Infections (C01) is dominant in intra-community connections (highest $s_c^{intra}$), while its role is less pronounced in the inter-community connection strength $s_c^{inter}$. This indicates that knowledge integration in infection-related medical research tends to reinforce existing linkages, which is confirmed by stronger connections between C01 terms in Fig. \ref{fig:1999I}. On the other hand, Nervous System Diseases (C10) play a much stronger role in inter-community connections by leading in the $s_c^{inter}$ ranking. This indicates that knowledge integration in nervous system diseases tends to bridge distant knowledge. This is confirmed in by observing the leading positions of C10 terms in top $b_i$ rankings in Table \ref{tab:local measures}. 

Furthermore, we note that Neoplasms (C04) manifests a relatively weaker intra-community connectivity (absence from top three $s_c^{intra}$) with a stronger inter-community connectivity (top two $s_c^{inter}$). Such a strong bridging role is again confirmed by the leading positions of C04 related terms in the $b_i$ rankings in Table \ref{tab:local measures}.
\begin{table}[ht!]
    \centering
    \begin{tabular}{cccc}
    \hline
      \textbf{Top} $\mathbf{s_i}$  & 1999 & 2010 & 2022 \\
    \hline
       I  & \Centerstack{C04.588(7.62)\\C04.557(6.52) \\ C12.050(5.95)} & \Centerstack{C04.588(7.84)\\C04.557(6.48) \\ C23.550(6.23)} & \Centerstack{C04.588(7.35)\\C04.557(5.95) \\  C12.050(5.67)} \\
    \hline
       NI June  &\Centerstack{C04.588(8.41)\\C04.557(7.58) \\ C12.050(6.35)} &\Centerstack{C04.588(8.22) \\ C04.557(6.99)\\C23.550(6.79)} &\Centerstack{C04.588(7.43) \\ C12.050(5.95) \\C16.131(5.94)} \\
    \hline
      \textbf{Top} $\mathbf{b_i}$  &  &  &  \\
    \hline
       I  & \Centerstack{C04.588(0.23)\\C10.228(0.17) \\ C16.131(0.16)} & \Centerstack{C04.588(0.29)\\C10.228(0.21) \\ C16.131(0.13)} & \Centerstack{C04.588(0.25)\\C10.228(0.15) \\ C10.551(0.11)} \\
    \hline
       NI June  & \Centerstack{C04.588(0.25)\\C10.228(0.18) \\ C10.551(0.14)} & \Centerstack{C04.588(0.29)\\C10.228(0.17) \\ C10.551(0.12)} & \Centerstack{C04.588(0.24)\\C10.228(0.16) \\ C16.131(0.15)} \\
       \hline
      \textbf{Top} $\mathbf{s_c}$   &  &  &  \\
    \hline
       I   & \Centerstack{C01(43.74)\\C10(39.76)\\C04(33.86)} & \Centerstack{C01(41.61)\\C10(37.22)\\C04(31.65)} & \Centerstack{C01(38.34)\\C10(38.01)\\C04(28.16)} \\
    \hline
       NI June   & \Centerstack{C01(44.54)\\C10(42.06)\\C04(34.28)} & \Centerstack{C01(43.45)\\C10(39.51)\\C04(32.82)} & \Centerstack{C01(39.42)\\C10(39.35)\\C04(27.18)} \\
       \hline
      \textbf{Top} $\mathbf{s_c^{intra}}$   &  &  &  \\
    \hline
       I   & \Centerstack{C01(20.32)\\C26(13.97) \\C10(12.15)} & \Centerstack{C01(19.80)\\C11(12.40) \\C26(11.93)} & \Centerstack{C01(16.55)\\C10(12.77) \\C26(11.93)} \\
    \hline
       NI June  & \Centerstack{C01(19.85)\\C10(13.24) \\C26(11.40)} & \Centerstack{C01(19.43)\\C26(12.65) \\C10(12.27)} & \Centerstack{C01(16.33)\\C26(14.98) \\C10(12.46)} \\
       \hline
      \textbf{Top} $\mathbf{s_c^{inter}}$   &  &  &  \\
    \hline
       I  & \Centerstack{C10(27.61)\\C04(25.29) \\C01(23.41)} & \Centerstack{C10(25.44)\\C04(24.19) \\C01(21.81)} & \Centerstack{C10(25.24)\\C04(22.18) \\C01(21.78)} \\
    \hline
       NI June  & \Centerstack{C10(28.82)\\C04(26.94) \\C01(24.69)} & \Centerstack{C10(27.25)\\C04(25.82) \\C01(24.01)} & \Centerstack{C10(26.89)\\C01(23.09) \\C04(21.81)} \\
       
  \hline
    \end{tabular}
    \caption{Top three second-level MeSH terms, ranked by node strength $s_i$ and betweenness $b_i$. The total connection strength $s_c$ of each parent category (first-level MeSH term) is computed by aggregating intra-category strength $s_c^{intra}$ and inter-category strength $s_c^{inter}$. Top three first-level categories with respect to $s_c$, $s_c^{intra}$ and $s_c^{inter}$ are listed. C04.588: Neoplasms by Site. C04.557: Neoplasms by Histologic Type. C12.050: Female Urogenital Diseases and Pregnancy Complication. C10.228: Central Nervous System Diseases. C23.550: Pathologic Processes. C16.131: Congenital Abnormalities. C10.551: Nervous System Neoplasms. C01: Infections. C10: Nervous System Diseases. C04: Neoplasms. C26: Wounds and Injuries. C11: Eye Disease. }
    \label{tab:local measures}
\end{table}

\subsection{Analyzing differences between I and NI-June Networks}
\label{sec: Differences between I and NI-June Networks}
In this section, we explore how I networks differ from NI-June networks in detail. Specifically, we study two questions: (i) do strong differences tend to be adjacent with strong differences? (ii) do differences cluster around certain topic areas? To this end, as described in Section \ref{Method}, we subtracted link weights of NI-June networks from I networks, i.e., $w_{ij}^I-w_{ij}^{NI-june}$, and constructed the positive and negative difference network separately for comparison.

In Section \ref{sec: Do strong differences co-locate?}, we aim to explore if strong differences co-locate. In Section \ref{sec: Do differences cluster?}, we aim to study the patterns of clustering and specific topics where positive and negative differences tend to be prominent.

\subsubsection{Do strong differences co-locate?}
\label{sec: Do strong differences co-locate?}
\begin{figure}
    \centering
    \includegraphics[width=0.8\linewidth]{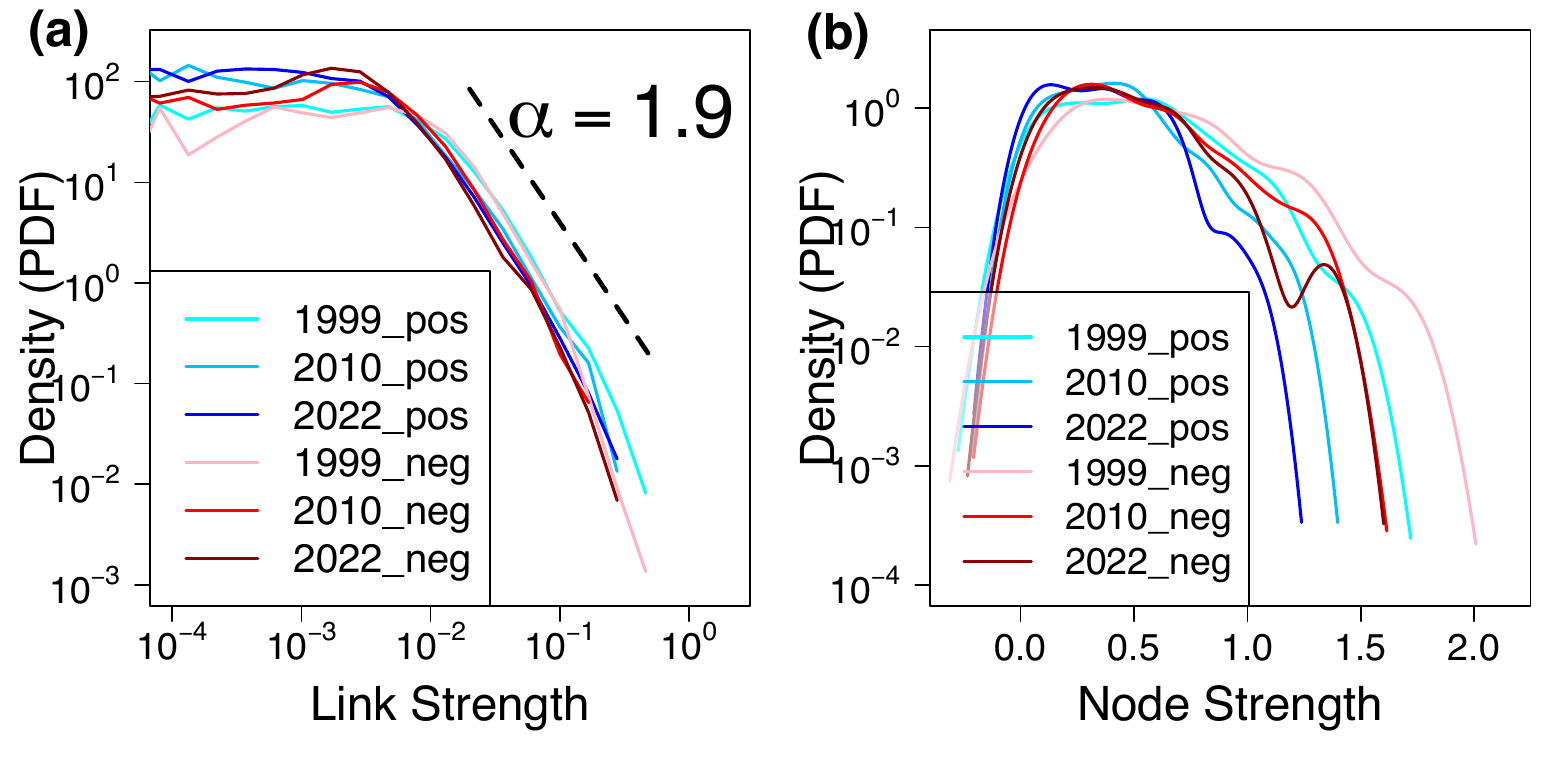}
    \caption{Link and node strength distribution for the difference networks. Left: link strength distribution with power law tail $\alpha=1.9$. Right: node strength distribution.}
    \label{diff dist}
\end{figure}

To gain intuition towards the overall connectivity of the constructed positive and difference network, we first plot the link strength distribution for positive and negative difference networks for 1999, 2010 and 2022 in Fig. \ref{diff dist}(a) and node strength distributions in Fig. \ref{diff dist}(b). We note that the distributions show approximately power-law distributions over several orders of magnitude in the x and y dimensions with a power law exponent close to $1.9$ in Fig. \ref{diff dist}(a).

\begin{figure}
    \centering
    \includegraphics[width=0.9\linewidth]{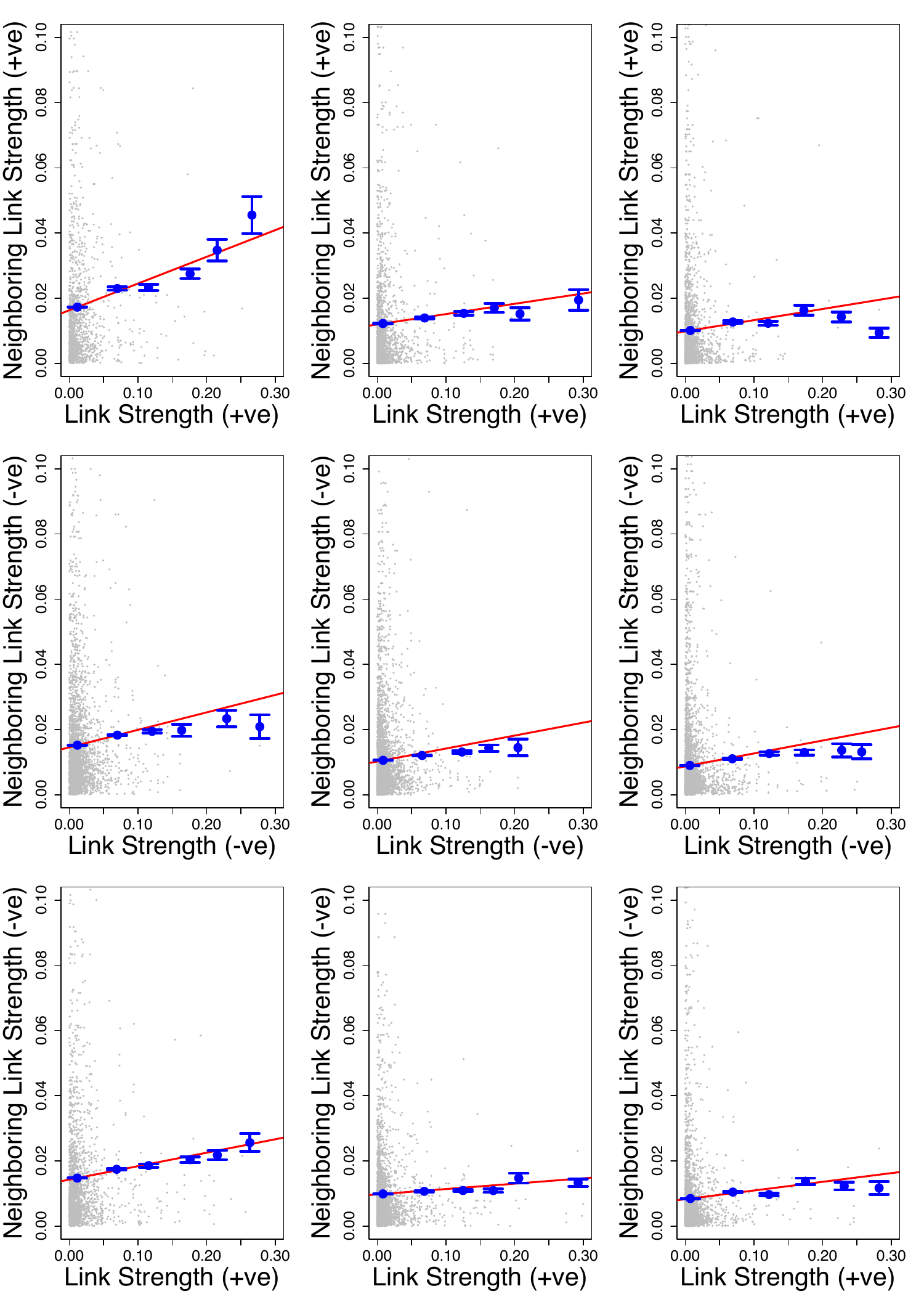}
    \caption{Plot of link co-location correlations for positive-positive (upper), negative-negative (middle), and positive-negative (lower) difference networks in 1999 (left), 2010 (middle), and 2022 (right). The x-axis represents the link strengths in the networks and the y-axis represents the link strengths of links adjacent to a given link on the x-axis. The red line is a fitted linear regression line. For positive-positive difference plots, the regression slopes are $0.08$ in 1999 ($R^2=0.007$), $0.03$ in 2010 ($R^2=0.001$) and $0.03$ in 2022 ($R^2=0.001$); while for negative-negative difference plots, $0.05$ in 1999 ($R^2=0.003$), $0.04$ in 2010 ($R^2=0.002$) and $0.04$ in 2022 ($R^2=0.002$), and for positive-negative difference plots, $0.04$ in 1999 ($R^2=0.004$), $0.02$ in 2010 ($R^2=0.001$), $0.03$ in 2022 ($R^2=0.001$). All six slopes reported are statistically significant ($p<2^{-16}$). Blue points represent the averages over bins, with error bars representing one standard error. We observe an overall weak tendency for larger differences to co-locate.}
    \label{fig: colocation}
\end{figure}

To explore the global patterns of connectivity of the positive and negative difference networks, we are particularly interested in the location of strong differences relative to each other. We first examine whether stronger links are more likely to be adjacent. To operationalise this idea, for the positive and negative difference networks, respectively, we consider the strength of all neighbouring links for each link and test whether stronger links tend to have stronger neighbouring links. Plotting link strengths on the x-axis and neighbouring link strength on the y-axis in Fig. \ref{fig: colocation}, we proceed with a correlation analysis and make the following observations. For positive-positive difference plots, the regression slopes are $0.08$ in 1999 ($R^2=0.007$), $0.03$ in 2010 ($R^2=0.001$) and $0.03$ in 2022 ($R^2=0.001$); while for negative-negative difference plots, $0.05$ in 1999 ($R^2=0.003$), $0.04$ in 2010 ($R^2=0.002$) and $0.04$ in 2022 ($R^2=0.002$), and for positive-negative difference plots, $0.04$ in 1999 ($R^2=0.004$), $0.02$ in 2010 ($R^2=0.001$), $0.03$ in 2022 ($R^2=0.001$).

Given the observed minor positive slopes, we conclude that there is a weak tendency for stronger links to pair with stronger neighbouring links in the difference networks. This indicates that topic-pairs where I (or NI-June) are relatively more prominent tend to be weakly adjacent. 

\subsubsection{Do differences between the networks cluster?}
\label{sec: Do differences cluster?}
\begin{figure}
    \centering
    \includegraphics[width=\linewidth]{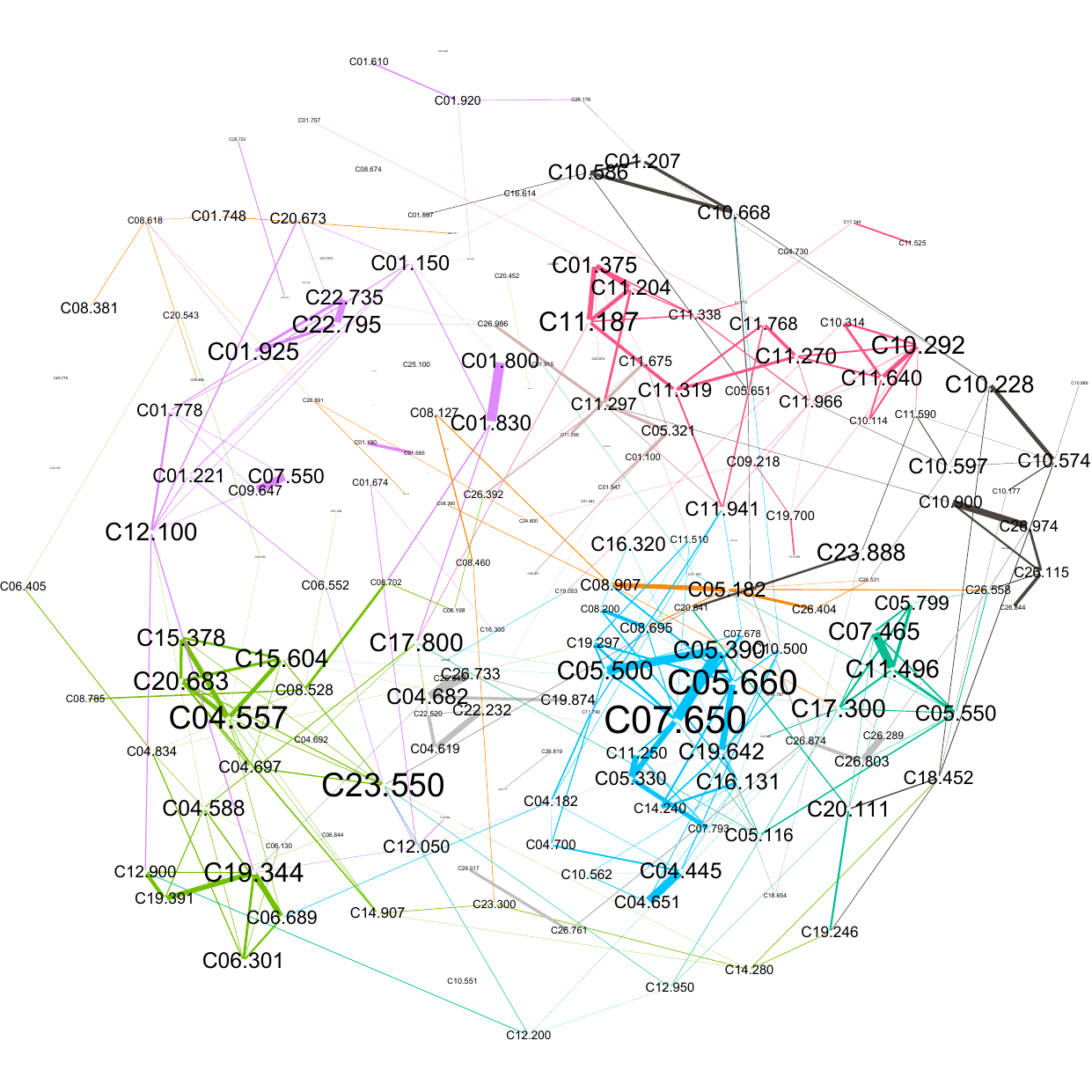}
    \caption{Positive difference network in 2022. Weaker links with weight less than $0.035$ were filtered out for clearer visualisation. Communities were detected based on the Louvain algorithm \cite{blondelFastUnfoldingCommunities2008} (modularity $Q=0.48$). Node label size indicates node strength. Link thickness indicates link weight.}
    \label{pos vis}
\end{figure}

\begin{figure}
    \centering
    \includegraphics[width=\linewidth]{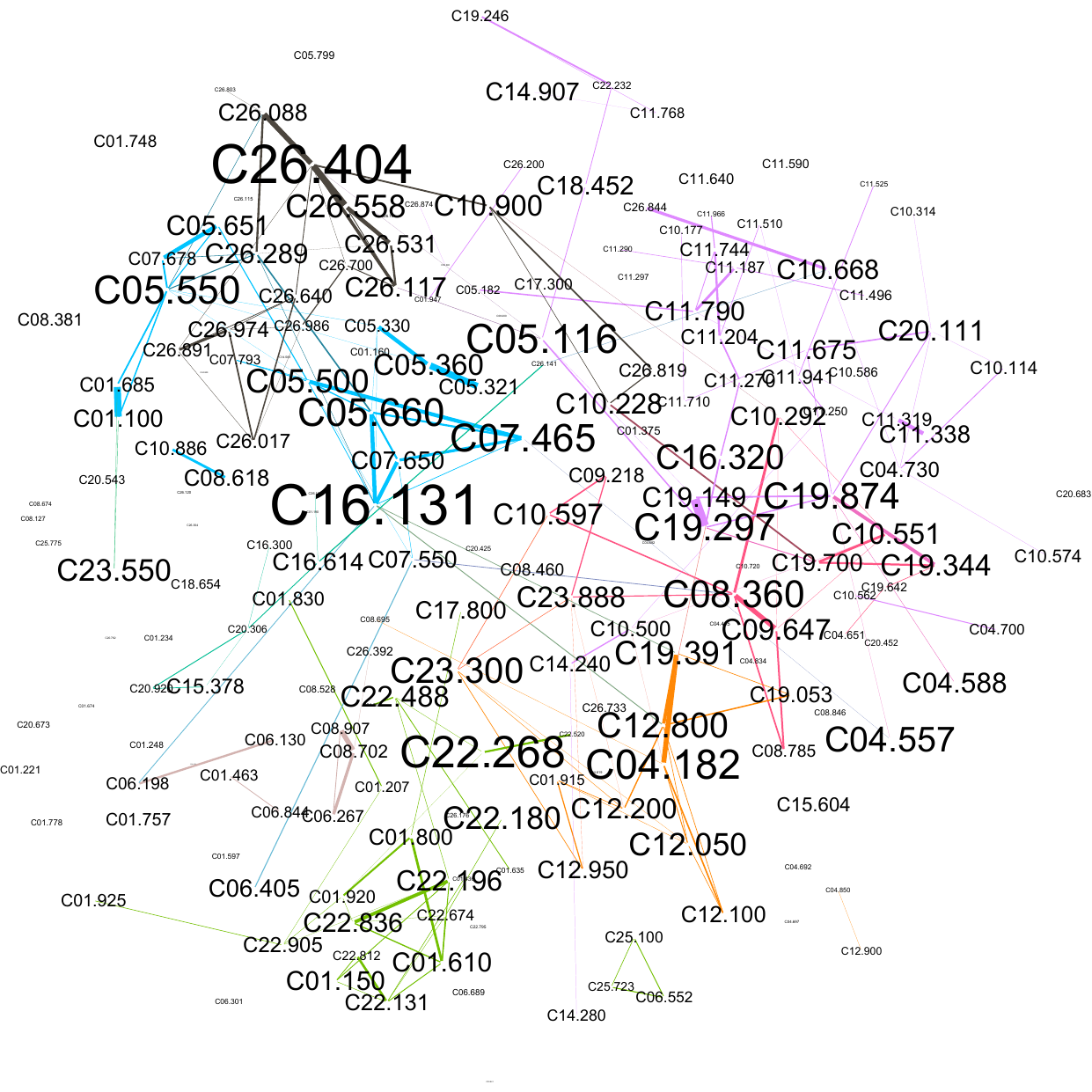}
    \caption{Negative difference network in 2022. Weaker links with weight less than $0.035$ were filtered out for clearer visualisation. Communities were detected based on the Louvain algorithm \cite{blondelFastUnfoldingCommunities2008} (modularity $Q=0.40$) and nodes were coloured accordingly. Node label size indicates node strength. Link thickness indicates link weight.}
    \label{neg vis}
\end{figure}

In Section \ref{sec: Do strong differences co-locate?}, we observed a weak tendency for differences to be co-located. Does this imply a potential clustering in the difference networks?
To gain intuition, we again first visualised positive and negative difference networks for the year 2022 (see Fig. \ref{pos vis} and \ref{neg vis}). Using the Louvain algorithm \cite{blondelFastUnfoldingCommunities2008}, we identified significant communities (modularity $Q=0.48$ in Fig.\ref{pos vis} and $Q=0.40$ in Fig.\ref{neg vis}), which corroborate that the topical differences tend to cluster.

Looking at the community membership for the positive difference network in Fig. \ref{pos vis}, there are four main clusters. The blue cluster consists of C07 (Stomatognathic Diseases) and C05 (Musculoskeletal Diseases) concepts. The warm green cluster consists of C04 (Neoplasms), C15 (Hemic and Lymphatic Diseases), and C19 (Endocrine System Diseases) concepts. The red cluster consists of many C11 concepts (Eye diseases). The purple cluster consists of multiple C01 (Infections) concepts.

In contrast, the community membership differs largely for the negative difference network in Fig. \ref{neg vis}. The blue cluster is led by C16 (Congenital, Hereditary, and Neonatal Diseases and Abnormalities) and C05 (Musculoskeletal Diseases) concepts. The black cluster is led by C26 (Wounds and Injuries) concepts. The green cluster is led by C22 (Animal diseases) and C01 (Infections) concepts. 

Interestingly, comparing above visualisations of differences with that of I and NI-June network ((Fig. \ref{fig:1999I}), we observed that there is certain degree of similarity of the leading nodes in each. Therefore, we hypothesise that stronger links in the original networks will tend to be stronger links in the difference networks. 

\begin{figure}
    \centering
    \includegraphics[width=0.8\linewidth]{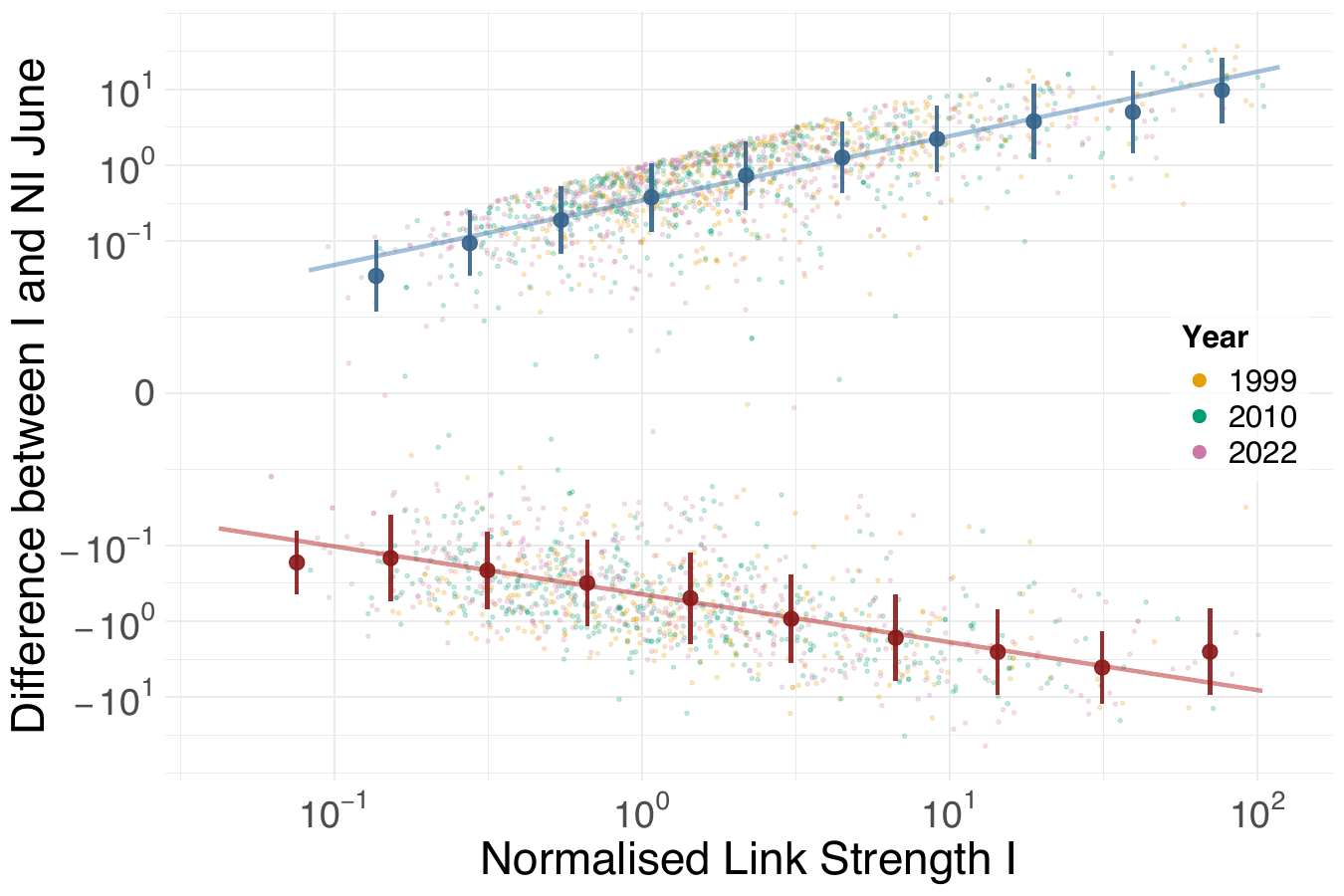}
    \caption{Dual power law of link differences (I-NI June) against link strengths of the I networks (1999, 2010, 2022), with logarithmic bins and error bars of one standard deviation for averages over the bins. The upper point cloud represents positive differences and closely matches a power law relationship with $\alpha=0.85$ (blue line) and negative $\alpha=0.63$ (red line). Both I and NI June networks have been normalised to the average link strength of the I networks as described in Section \ref{Method}. }
    \label{fig: diff vs I}
\end{figure}

To test this hypothesis, we consider the link strength of the I networks of 1999, 2010 and 2022, and examine whether the corresponding differences (I-NI June) tend to be stronger for stronger I network links. We plot link strengths of I networks of the three years on the x-axis and the corresponding differences on the y-axis in Fig. \ref{fig: diff vs I}. In addition, to indicate estimations for potential power-law relationships, we plot linear regression lines of best fit to the positive and negative point clouds, respectively. We make the following observations. 

Both positive and negative link differences follow a power-law relationship with respect to the link strength of I networks, with the power law exponent $\alpha=0.85$ ($R^2=0.53,p<0.001$) for the positive differences and $\alpha=0.63$ ($R^2=0.30,p<0.001$) for the negative differences. This indicates that it is true that stronger links in I networks tend to be stronger links in the difference networks; moreover, both the positive and negative differences follow a power-law scaling with the original link strengths of I networks. We note that the cut-off pattern observed at the upper point cloud is the upper-bound of the link differences, i.e., link strengths of the I networks.

\begin{figure}
    \centering
    \includegraphics[width=0.95\linewidth]{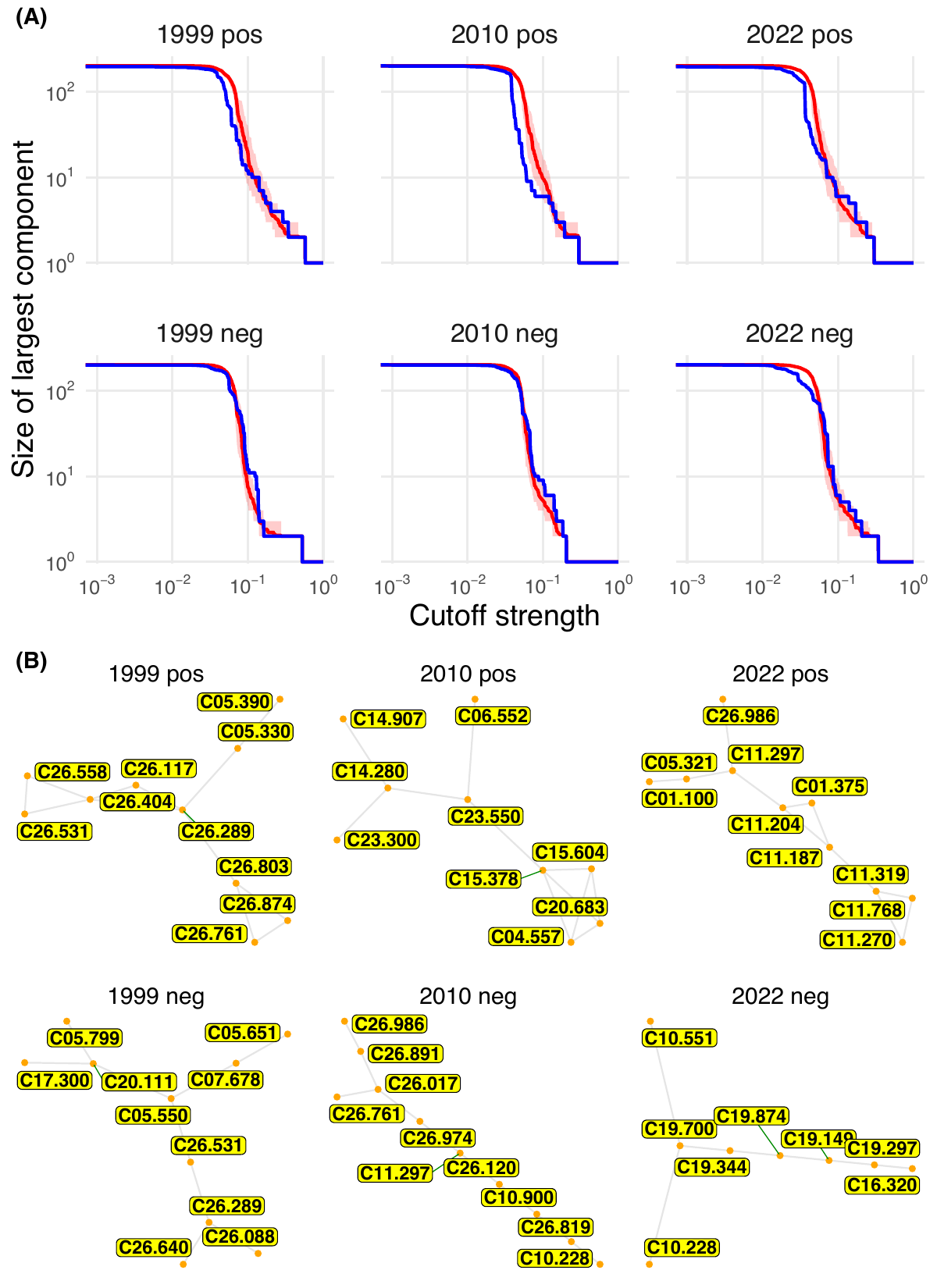}
    \caption{(A) Dependence of the size of the largest connected component (LCC) on the cutoff threshold for positive and negative difference networks for all three points in time. The blue line represents the empirical network and the red line represents the mean of 100 randomised networks (through random link shuffling) with 95\% confidence interval (red ribbon). (B) Inner core component membership (as defined in Fig.\ref{LCC}(B)) for positive and negative difference networks. There are cases where LCC of size 10 does not exist. In those cases LCC of size that are immediately smaller than 10 are shown, namely, size 9 for 2010 pos, size 8 for 2022 neg, and size 10 for 1999 pos, 2022 pos, 1999 neg, 2010 neg.}
    \label{core pos neg}
\end{figure}

To better understand the link arrangement in the positive and negative difference networks and the areas of relative prominence, we repeat the core analysis for the difference networks in Section \ref{sec: Core Analysis} by visualising their LCC-cutoff dependency in Fig.\ref{core pos neg}(A) and inner core in Fig.\ref{core pos neg}(B).

Analysing Fig.\ref{core pos neg}(A), we observe that the LCC decomposition curve (blue) of all positive difference networks and negative difference network in 2022 exhibited an initially slightly faster decay than expected at random (red) for low cut-off thresholds around 0.01-0.1. This again indicates that these difference networks have core-periphery structures. In contrast, the decompositions of negative difference networks in 1999 and 2010 mostly coincide with what would be expected at random.

We next identify the areas with prominent differences between the I and NI networks through the visualisation of the inner cores of the positive and negative difference networks in Fig.\ref{core pos neg}(B).  The core of the 1999 positive difference network builds around many C26 concepts (Wounds and Injuries), whereas in 2010 it becomes a mixture of C23 (Pathological Conditions, Signs and Symptoms), C15 (Hemic and Lymphatic Diseases), and C14 (Cardiovascular Diseases) concepts and in 2022 a focus on C11 concepts (Eye Diseases). For the negative difference networks, the core in 1999 builds around C05 (Musculoskeletal Diseases) and C26 (Wounds and Injuries) concepts, yet changes to C10 (Nervous System Diseases) and C26 (Wounds and Injuries) concepts in 2010, and finally changed to C19 (Endocrine System Diseases) concepts.

We conclude that there is little agreement on the inner core memberships (i) between positive and negative networks, and (ii) across time. The discrepancy for (i) indicates different areas of relative prominence for I and NI networks, which may originate from the difference in the relative focus and research practices in the impactful or less impactful journals.

For (ii), the varying core memberships indicates the relative prominence of impactful or less impactful IDR changes through decades, where domain trends, research practices, and external events like pandemics might be contributing factors. Such temporal heterogeneity of relative prominence provides a strong contrast to the absolute prominence observed earlier in the core membership in Fig. \ref{LCC}(B). This implies that despite the time-variant areas of relative prominence of impactful or less impactful IDR across decades, there is a fundamentally universal knowledge structure in medicine that serves as a stable core across decades.

\section{Conclusion}
In this paper, we characterised interdisciplinary knowledge structure of PubMed research articles in medicine as correlation networks of medical concepts and compared the interdisciplinarity of articles between highly impactful medical journals and others. We draw the following conclusions to answer the research questions we started with.

We found that highly impactful medical journals tend to produce less interdisciplinary research than less impactful journals. We also established that, as it bridges distant knowledge clusters, cancer-related research can be seen as the main driver of interdisciplinarity in medical science. 

To explore if the differences between impactful and less impactful networks cluster, we subtracted the less impactful network from the impactful network and compared the resulting positive and negative difference networks. We first concluded that, in terms of link arrangement, there is a weak tendency for stronger links to co-locate in the difference networks. We then established that the differences tend to cluster around certain areas that change dramatically over time, representing evolving topics of relative prominence of impactful and less impactful journals. In contrast, both impactful and less impactful journals agree on the areas of absolute prominence, which can be seen as a universal core knowledge structure in medicine that remains stable across decades.

Despite IDR being crucial for innovation and addressing complex societal issues, our results suggest more impactful medical journals that shape domain trends and practices tend to publish relatively more disciplinary-based articles. One potential contributing factor could be that peer-reviewers with bounded rationality tend to penalise innovative work \cite{boudreau_looking_2016,packalen_nih_2020,woelertParadoxInterdisciplinarityAustralian2013,duIntegrationVsSegregation2025a}. 

Another, or a perhaps more fundamental reason, comes down to how medical systems enact change. High-impact medical journals aim to publish studies that can be translated rapidly into clinical guidance; guideline developers such as National Institute for Health and Care Excellence (NICE)\footnote{\url{https://www.nice.org.uk/process/pmg20/chapter/reviewing-evidence}} give randomised evidence and well-conducted systematic reviews the highest initial certainty (according to GRADE\footnote{\url{https://www.cdc.gov/acip-grade-handbook/hcp/chapter-7-grade-criteria-determining-certainty-of-evidence/index.html}}), making them the most immediately actionable. Many IDR or complex interventions tends to be frequently downgraded for inconsistency, performance bias, and study design \cite{movsisyanOutcomesSystematicReviews2016}. 

In parallel, journals enforce prospective trial registration and standardised reporting (CONSORT 2025\footnote{\url{https://www.equator-network.org/reporting-guidelines/consort/}} for randomised trials; PRISMA 2020\footnote{\url{https://www.prisma-statement.org}} for systematic reviews), which favour tight pre-specified questions, single primary outcomes, and clean comparators. This compliance infrastructure makes discipline-anchored studies the path of least resistance, while IDR, with multiple components and mixed outcomes, is harder to pre-specify and to report crisply. Taken together, these forces impose a structural disadvantage on IDR in today’s evidence-to-guideline pipeline.


Loosening rigid disciplinary boundaries at elite journals \cite{du_prestigious_2025_frccs} doesn’t fully solve the problem because editorial strategy is instrumentally tuned to how evidence is operationalised in regulation and guidance. A perhaps more scalable fix is to accommodate IDR evaluation within the existing compliance infrastructures, with a positive example being the recent joint endorsement of the CONSORT-AI reporting extensions from leading medical journals \cite{GuidingBetterDesign2020}. This enables journals to capture strengths from other fields without sacrificing the clarity needed for practice change—ultimately providing the best opportunities to improve population health.




\newpage
\backmatter








\section*{Declarations}

\subsection*{Funding}
No funding was received for this study.
\subsection*{Acknowledgement}
Not applicable.
\subsection*{Competing interests}
The authors declare no competing interests.
\subsection*{Data and code availability}
Data and code is available on reasonable request.
\subsection*{Author contribution}
\textbf{Anbang Du:} Conceptualization, Software, Data curation, Resources, Formal Analysis, Methodology, Visualization, Writing – original draft,  Writing – review \& editing

\textbf{Michael Head:} Supervision, Data curation, Resources, Writing – review \& editing

\textbf{Markus Brede:} Conceptualization, Supervision, Methodology, Writing – review \& editing

\noindent

\end{document}